\def\year{2020}\relax
\documentclass[letterpaper]{article} 

\usepackage{booktabs} 
\usepackage{amsmath} 
\usepackage{subfigure} 
\usepackage{cite}
\usepackage{makecell}

\usepackage{aaai20}  
\usepackage{times}  
\usepackage{helvet} 
\usepackage{courier}  
\usepackage[hyphens]{url}  
\usepackage{graphicx} 
\usepackage{graphicx}  
\title{Adabot: Fault-Tolerant Java Decompiler}
\author{Zhiming Li\textsuperscript{\rm 1\thanks{Zhiming Li, Qing Wu and Kun Qian are co-first authors.}},Qing Wu\textsuperscript{\rm 1*},Kun Qian\textsuperscript{\rm 2*} \\ 
China University of Geosciences,Wuhan 430074,China\\
\textsuperscript{\rm 1}\{zhimingli, iwuqing\}@cug.edu.cn; \textsuperscript{\rm 2}kun\_qian@foxmail.com
}

\begin{document}
\maketitle
\begin{abstract}
Reverse Engineering(RE) has been a fundamental task in software engineering. However, most of the traditional Java reverse engineering tools are strictly rule defined, thus are not fault-tolerant, which pose serious problem when noise and interference were introduced into the system. In this paper, we view reverse engineering as a statistical, machine translation task instead of rule-based task, and propose a fault-tolerant Java decompiler based on machine translation models. Our model is based on attention-based Neural Machine Translation (NMT) and Transformer architectures. First, we measure the translation quality on both the redundant and purified datasets. Next, we evaluate the fault-tolerance (anti-noise ability) of our framework on test sets with different unit error probability (UEP). In addition, we compare the suitability of different word segmentation algorithms for decompilation task. Experimental results demonstrate that our model is more robust and fault-tolerant compared to traditional Abstract Syntax Tree (AST) based decompilers. Specifically, in terms of BLEU-4 and word error rate (WER), our performance has reached 94.50\% and 2.65\% on the redundant test set; 92.30\% and 3.48\% on the purified test set.
\end{abstract}
\section{Introduction}
Reverse Engineering has been an extremely important field in software engineering, it helps us to better understand the internal architecture and interrealtions of binary applications. Classical Java reverse engineering task includes disassembly and decompilation. Concretely, disassembly means mapping executable bytecode into mnemonic text representations. Whereas decompilation means mapping bytecode or the disassembled mnemonic text representations into reader-friendly source code. Though the procedure seems ideal and straight-forward, it is essentially an extremely difficult and routine task that requires mentally mapping assembly instructions or bytecode into higher level abstractions and concepts. Moreover, traditional disassemblers and decompilers are strictly rule defined that anything nonconforming would be spit out as error message, which is common since the source bytecode is usually informal and sometimes deliberately obfuscated for safety concern. These external obfuscations could be considered as noise and interference. Therefore, we conclude that the traditional rule-based decompilers are not fault-tolerant (anti-noise).

There are many recent works on computational linguistics of computer languages: mapping natural language(NL) utterances into meaning representations (MRs) of source code based on Abstract Syntax Tree (AST) \cite{yin2018tranx}, generating code comments for Java methods based on NMT and AST \cite{hu2018deep}, predicting procedure names in stripped executables \cite{david2019neural}, etc. It's obvious from the above mentioned that previous works focus on either code generation from natural language, or extraction of lexical information from the mnemonic, regardless of the structural information in decompiled source code.

Therefore, in order to build a functional fault-tolerant decompiler, we need to take both the lexical and structural (syntactic) information into consideration. Specifically, we need to combine the lexical information extracted from the bytecode or mnemonic (usually 1-to-1) with corresponding structural information to form syntactically readable source code.

\begin{figure*}[t]
\centering
\includegraphics[width=0.9\textwidth,clip=true]{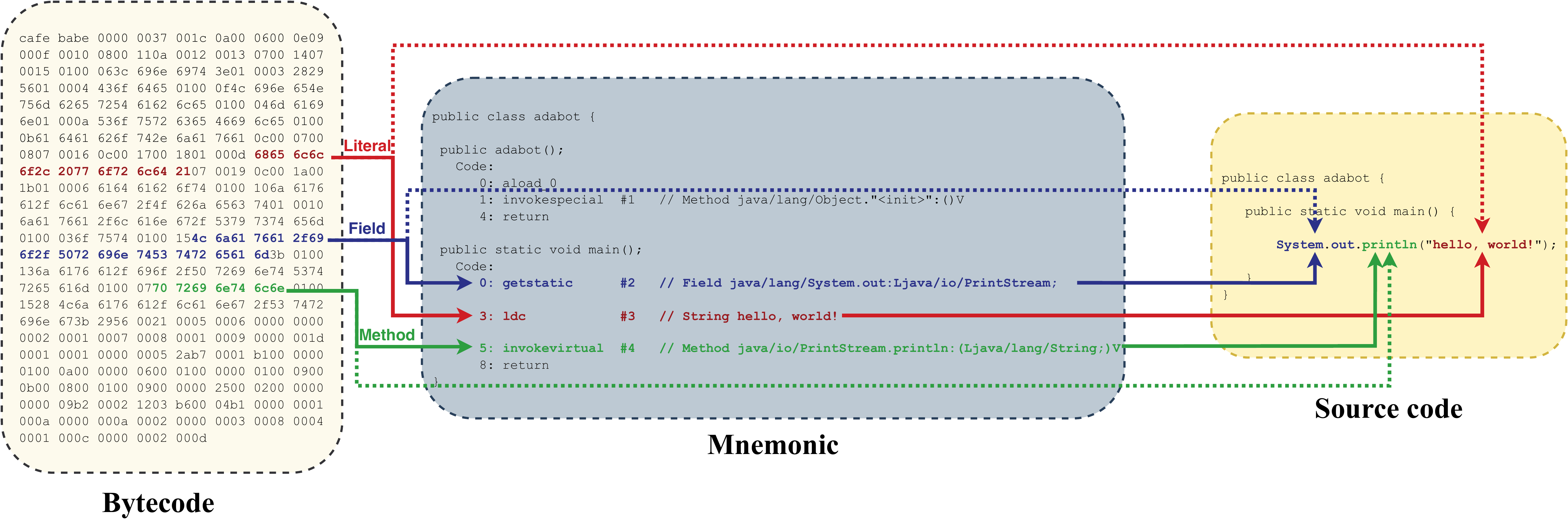} 
\caption{Instance of a parallel bytecode, mnemonic and source code triple. Literal, Field and Method indicate the lexicon ought to be extracted from bytecode or mnemonic by decompiler. While the corresponding source code indicates the relation between lexical and structural information decompiler should learn.}
\label{fig1}
\end{figure*}

Figure \ref{fig1} shows a concrete example of parallel bytecode, mnemonic and source code triple. Intuitively, decompiler extracts lexical information from bytecode or mnemonic and uses it to construct a corresponding source code. The closed loops in Figure \ref{fig1} illustrates the relationship between these three files in the process of decompilation. Concretely, We can either decompile from the soild arrow \cite{he2018debin,david2019neural}, which is from bytecode to mnemonic, then to source code. Or we can decompile from the dashed arrow, which is directly from bytecode to source code. However, both bytecode and mnemonic suffer from large redundancy and unbalanced distrbution of information compared with source code. Thus it requires our model capable of handling these problems in order to be functional and robust.

Although decompilation of programming language is highly similar to machine translation of natural language, they are actually different in many ways, which is as:

\begin{enumerate}
\item \textbf{Syntax Structure:} Programming language is rigorously structured \cite{hellendoorn2017deep,hu2018deep} that any error in source code is significant enough to make an AST based decompiler invalid.
\item \textbf{Vocabulary Distribution:} The frequency distribution of words in vocabulary of programming language is large in variance. Specifically, the vocabulary consists of unique and rare identifiers that are evenly frequent in use. This difference is concretely discussed in the following section from the perspective of information theory.
\item \textbf{Word Unit:} Programming language is less likely to suffer from out-of-vocabulary problem. Unlike natural language that can be variably expressed (e.g. blackbox = black-box = black box), words in the vocabulary of programming language are rigorously definfed. Therefore, subword-unit based word segmentation, which works great in handling out-of-vocabulary problem of natural language is not necessary for programming language for its vocabulary can be exhaustively learned.
\end{enumerate}
Here we analyze the second difference from the perspective of information theory.
\begin{table}[h]
	\centering
	\caption{Comparison of languages in terms of entropy and redundancy}\smallskip
\resizebox{.95\columnwidth}{!}{
\smallskip
	\begin{tabular}{ccc}
		\toprule  
		Language&Entropy (bit)&Redundancy \\ 
		\midrule  
		Natural Language (English)&11.82&0.09 \\
		Source Code&6.41&0.40 \\
		Bytecode&5.31&0.28 \\
		\bottomrule  
	\end{tabular}
}
\label{table1}
\end{table}

As Tabel \ref{table1} illustrated, we assess printed English, Java source code and bytecode of our dataset in terms of entropy and redundancy. Compared with printed English \cite{shannon1951prediction}, source code and bytecode are much lower in entropy and higher in redundancy. Furthermore, they have an extremely unbalanced distribution of information in vocabulary. Concretely, structural information like keywords, which only accounted for 0.4\% of the vocabulary, significantly contributed about 9\% to the overall redundancy.

Therefore, in order to properly handle the unbalanced distribution of information in source code and bytecode, our model should incorporate not only the ability of learning structural information, but also the ability of extracting lexical information out of redundancy. In our work, we first assess these two abilities of attention-based NMT and Transformer architectures \cite{vaswani2017attention} without any manual operation. Then we attach a purification operation that manually extracts lexical information of the identifiers from the structural information in order to further boost the performance of both architectures on decompilation task.

We evaluate the performance of our model on compilable snippets of all official Java 11 API offered by Oracle\footnote{https://docs.oracle.com/en/java/javase/11/docs/api/index.html}. Experimental results demonstrate that our model is capable of performing both high-quality and fault-tolerant decompilation. To our knowledge, this is the first work on observing the ability of machine translation models for fault-tolerant decompilation task. 

The contributions of this paper can be concluded into the followings:
\begin{itemize}
\item We propose a statistical, fault-tolerant Java decompiler.
\item We evaluate different word segmentation algorithms for programming language and conclude which one is the most appropriate.
\item We propose word error rate (WER) as a more reasonable metric for the evalauton of programming language than BLEU-4 \cite{papineni2002bleu}.
\item We demonstrate that Transformer is better in handling the unbalanced distribution of information in programming language than attention-based NMT.
\end{itemize}

\begin{figure*}[t]
\centering
\includegraphics[width=0.9\textwidth,clip=true]{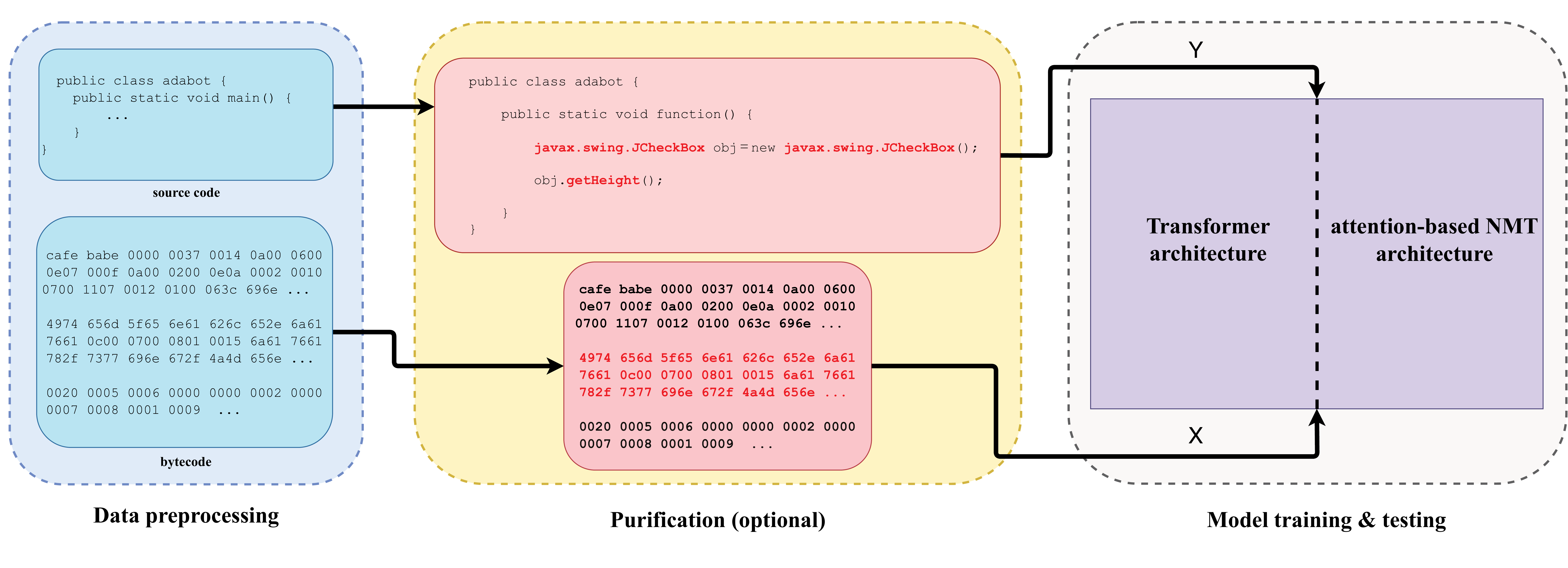} 
\caption{Overall workflow of Adabot}
\label{fig2}
\end{figure*}

\section{Approach}

From the analysis above, unlike natural language translation, programming language decompilation requires the model capable of properly handling the unbalanced distribution of information. Concretely, to learn not only the structural information consists of keywords and operators that significantly contributes to redundancy, but also the lexical information of identifiers that are comparitively high in entropy. Intuitively, it's not hard for the model to grasp the overall structural information since it's high in redundancy and appears repetitively in all the samples. However, it pose greater challenge for the model to learn the lexical information from high redundancy.

To address this problem, we propose taking use of the recurrence and attention mechanism in attention-based NMT and Transformer architectures. The overall workflow of Adabot is illustrated in Figure \ref{fig2}. It mainly consists of three parts: data preprocessing, purification (optional), model training and testing. To obtain the dataset, we first crawl all the official Java 11 API offered by Oracle. Then reflect those available from the local packages with Java reflection mechanism. Finally we compile them into corresponding bytecode with format templates. The details of data preprocessing can be found in the Experiments section. In our work, we find that Transformer architecture which is solely based on attention mechanism, is capable of learning not only the structural information (black characters in the second dotted box) but also the lexical information(red characters in the second dotted box). Whereas, attention-based NMT architecture is only capable of learning the structural information but not the lexical information without manual assistance. It is because attention-based NMT architecture is essentially based on recurrence mechanism and only used attention mechanism as an auxiliary operation to relieve the vanishing gradient problem which is still inefficient when dealing with long sequences. Therefore, this halfway solution is invalid for the bytecode in our dataset since the average sequence length is about 400. In order to fully boost the potential of the model, we attach a manual purification step for attention-based NMT before training. Specifically, this step helps to acquire the purified dataset by removing a large proportion of structural information in both bytecode and source code, which significantly alleviate the vanishing gradient problem.

\subsection{Word Segmentation}

Machine Translation of natural language is an open-vocabulary problem because of its variability in expression, such as compunding (e.g., blackbox = black-box = black box), morpheme (e.g., likely and un-likely), etc. Therefore, it's hard for machine to tell whether it is a compound phrase or a single word that hasn't been learned before, which face the model with serious out-of-vocabulary problem. Popular word segmentation algorithms that address this problem includes back-off dictionary \cite{jean2014using,luong2015effective} and subword model based on byte-pair-encoding (BPE) \cite{gage1994new} algorithm \cite{sennrich2015neural}.

\begin{figure}[t]
\centering
\includegraphics[scale=0.54,clip=true]{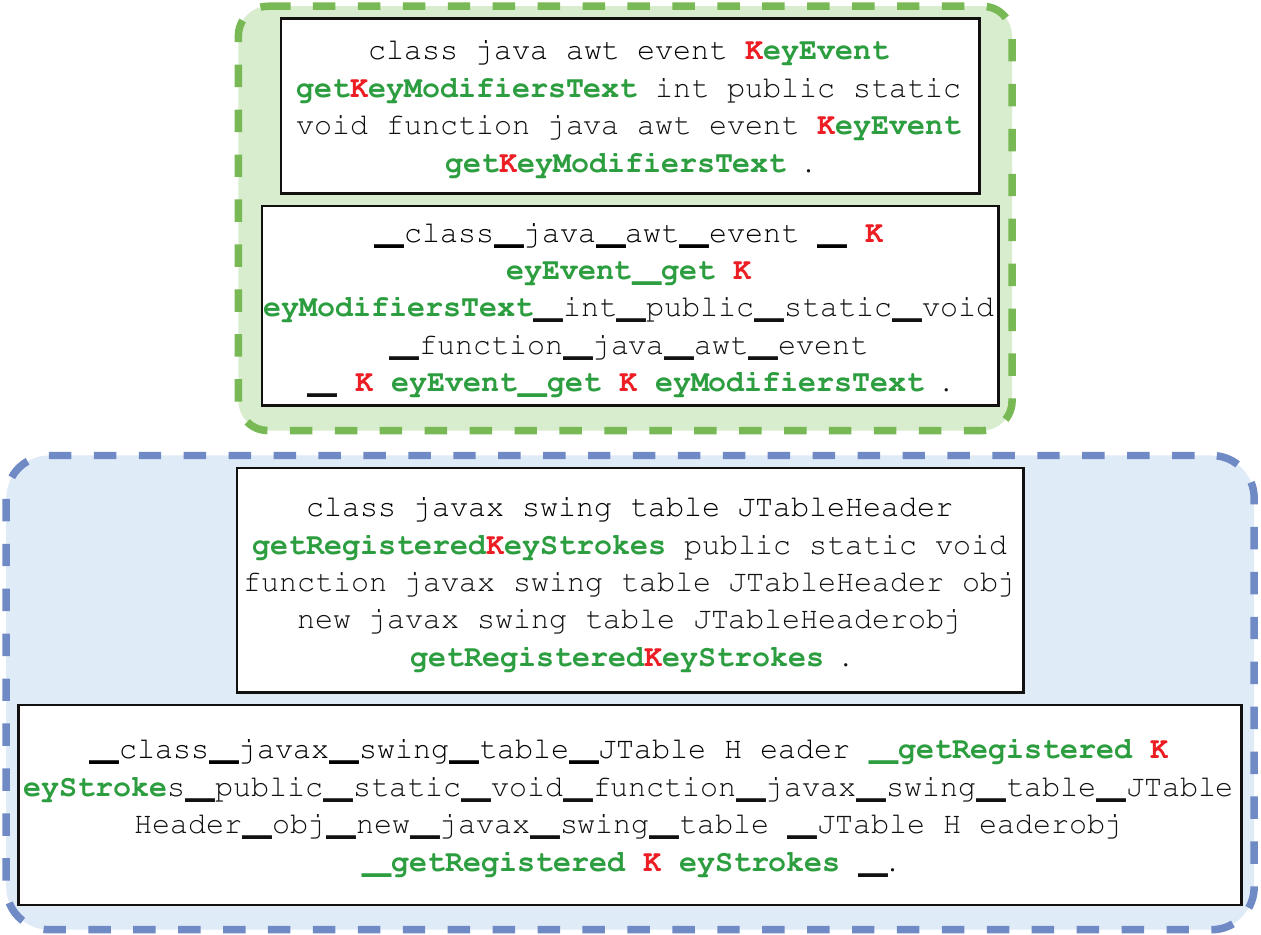} 
\caption{An example of BPE based word segmentation}
\label{fig3}
\end{figure}

In our work, we evaluate the adaptability of different word segmentation algorithms on the decompilation task, including space delimiter and subword model based on BPE. Space delimiter is self-explanatory, simply use space as the word delimiter. The following introduces the basic concept of subword model based on BPE. Intuitively, the motivation of subword segmentation is to make compounds and cogantes transparent to machine translation models even if they haven't been seen before. Specifically, BPE based word segmentation initializes with representing each word in the vocabulary as a sequence of characters and iteratively merge them into n-gram symbols. Figure \ref{fig3} is an example of BPE based word segmentation on our dataset. Specifically, the vocabulary of programming language is relatively small and the words in it are all case sensitive. Therefore, the left and right sequences of character ``K'' appear differently every time. As a result, ``K'' is remained as an independent symbol in the vocabulary. This cause serious problem for it compromises the meaning of identifier(``KeyEvent'', ``getKeyModifiersText'', etc.) and make it hard for the model to understand it as an entity.

\subsection{Attention in Transformer}

Self-attention mechanism is what actually makes Transformer so powerful in handling the unbalanced distribution of information. The architecture is entirely based on self-attention mechanism instead of recurrence which thoroughly resolves the vanishing gradient problem found in NMT. Concretely, the model first assigns three vectors for each of the input words, including query vector $Q$, key vector $K$ and value vector $V$. Then, the attention score of each word is calculated with $Q$ and $K$ and passed through the softmax layer to produce the final attention weight. Eventually, the attention weight and $V$ of every words in the input sequence are used to get the output attention vector based on the scaled dot-product attention function:

\begin{equation}
\text { Attention }(Q, K, V)=\operatorname{softmax}\left(\frac{Q K^{T}}{\sqrt{d_{k}}}\right) V
\end{equation}

\begin{figure}[h]
\centering
\includegraphics[scale=0.3,clip=true]{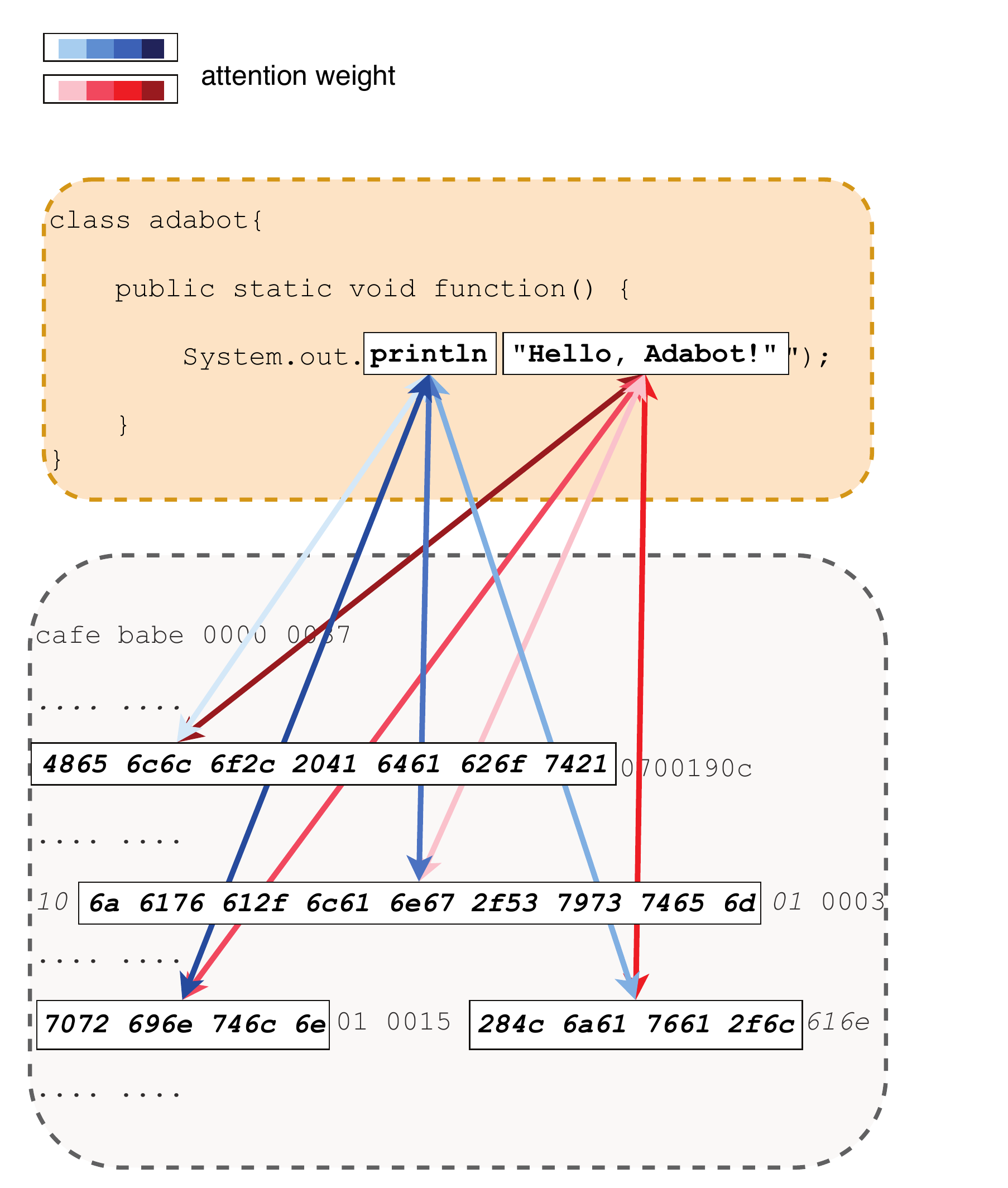} 
\caption{Visualization of attention}
\label{fig4}
\end{figure}

Figure \ref{fig4} is a specific example of the attention mechanism. The different shades of color indicates the significance of weighted attention. For example, the String ``Hello, Adabot!'' in source code has stronger attention with bytecode units that represent ``Hello, Adabot!'', ``java/Lang/String'', ``println'' (in descending order) than other irrelevant units.

In conclusion, being entirely based on self-attention mechanism allows Transformer to process all the bytecode units in parallel before deciding which of those deserve more attention. Thus makes it better in handling the unbalanced distribution of information.

\subsection{Attention in NMT}

\begin{figure}[t]
\centering
\includegraphics[scale=0.6,clip=true]{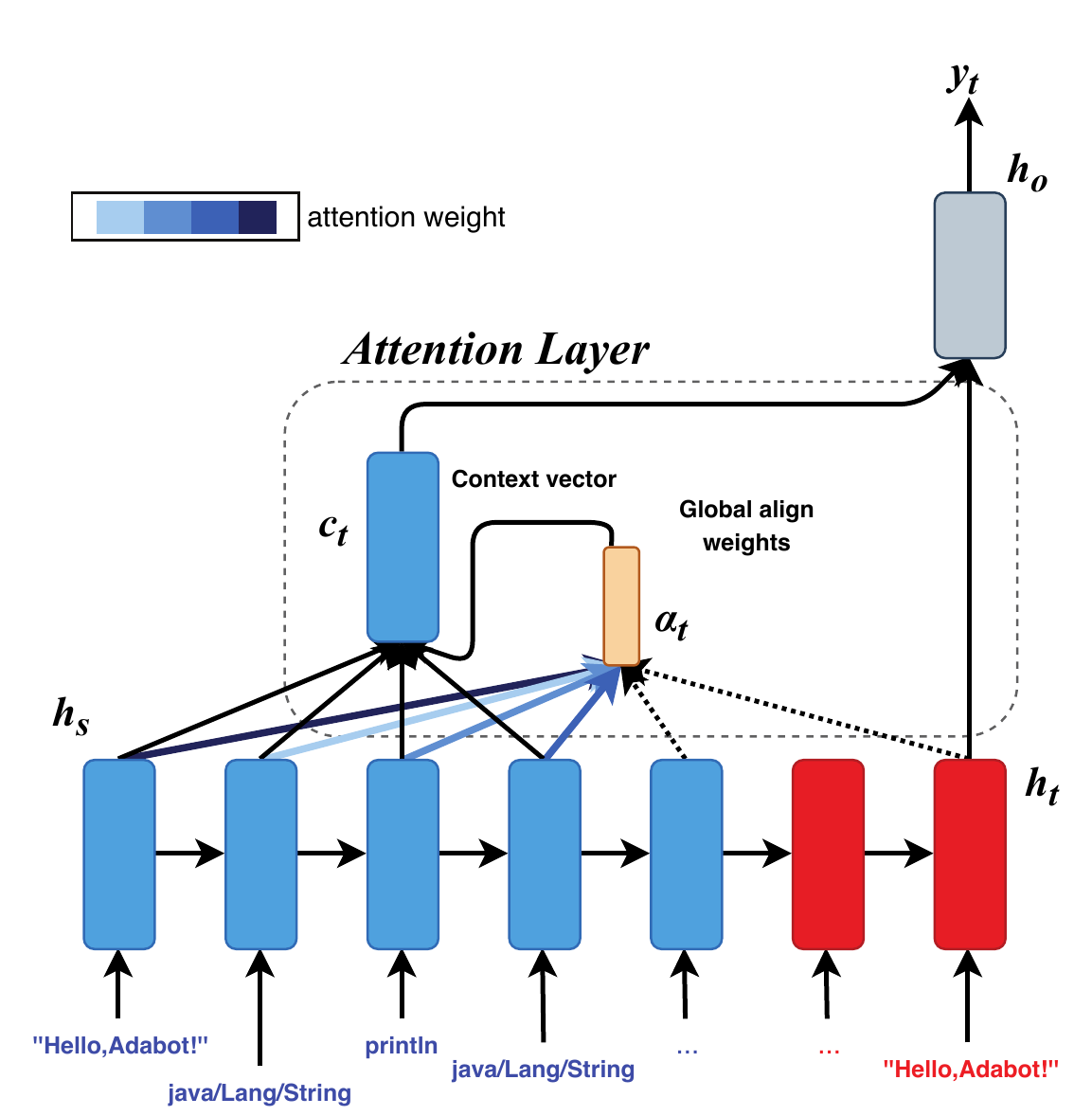} 
\caption{Attention mechanism in NMT}
\label{fig5}
\end{figure}

Attention mechanism in NMT allows the model to pay attention to relevant source content during translation based on the short-cut alignment between the source and the target. Intuitively, this helps alleviate the potential vanishing gradient problem caused by the distance between relevant source units and target word. The global attention-based NMT model is illustrated in Figure \ref{fig5}. Blue arrows with different color shades indicate the significance of attention weights. Concretely, the model first compare the target hidden state $h_{t}$ with all the source states $h_{s}$ to get the corresponding attention weights $\alpha_{ts}$, which is as follows:

\begin{equation}
\begin{split}
\alpha_{t s} &=\operatorname{align}\left(\boldsymbol{h}_{t}, \boldsymbol{h}_{s}\right) \\ &=\frac{\exp \left(\operatorname{score}\left(\boldsymbol{h}_{t}, \boldsymbol{h}_{s}\right)\right)}{\sum_{s^{\prime}} \exp \left(\operatorname{score}\left(\boldsymbol{h}_{t}, \boldsymbol{h}_{s^{\prime}}\right)\right)}
\end{split}
\end{equation}

Then $\alpha_{ts}$ is used to get the weighted context vector $c_{t}$, which is:

\begin{equation}
\boldsymbol{c}_{t}=\sum_{s} \alpha_{t s} \boldsymbol{h}_{s}
\end{equation}

Finally, the model combines $c_{t}$ with the current target state to get the attention vector which is used as the output prediction $h_{o}$ as well as the input feeding for the next target $h_{t+1}$, the attention vector is computed as follows:

\begin{equation}
\boldsymbol{a}_{t}=f\left(\boldsymbol{c}_{t}, \boldsymbol{h}_{t}\right)=\tanh \left(\boldsymbol{W}_{\boldsymbol{c}}\left[\boldsymbol{c}_{t} ; \boldsymbol{h}_{t}\right]\right)
\end{equation}

Intuitively, we can see that the global attention mechanism allows NMT to focus on the relavant bytecode units that represent ``Hello, Adabot!'', ``java/Lang/String'', ``println'', etc. (in descending order) which are distant from the current target source code ``Hello, Adabot!''. However, serving as an auxiliary mechanism aiming to alleviate the vanishing gradient problem, global attention in NMT alone is not significant enough to handle the unbalanced distribution of information in programming languages. Therefore, manual operation which we called purification is required.

\subsection{Introduction of Noise}

In our work, in order to evaluate the fault-tolerance (anti-noise ability) of our model, we introduce noise in the form of salt-and-pepper noise (a.k.a. impulse noise). Specifically, each unit in the source bytecode shares a probability of $p_{u}$ being corrupted into either 0xff (salt) or 0x00 (pepper) .

The bit error probability (BEP) is a concept used in digital transmission, which is the prior probability of a bit being erroneous considering each bit as an independent variable. It is used as an approximate estimation of the actual bit error rate (BER). Here we introduce the concept of unit error probability (UEP) which is similar to BEP but takes two bytes as one basic unit instead of one bit since the Java virtual machine takes two bytes as one basic unit. It is used as an approximate estimation of the actual unit error rate (UER), which is computed as follows:

\begin{equation}
UER \approx p_{u} N
\end{equation}
where  $p_{u}$ is unit error probability, $N$ is the number of units in one bytecode sample.

\subsection{Evaluation Metric of Reverse Engineering}

So far there has not been an official measure for the evaluation of code generation task (e.g. reverse engineering, program synthesis, etc). Popular measures implemented by previous work includes: BLEU-4 which has been exploited for the evaluation of API sequences generation as well as comment generation for Java methods \cite{gu2016deep,hu2018deep}; exact match (EM) and execution (EX) accuracy which has been used for the evalution of code generation from queries \cite{yin2018tranx}. However, from our experiment, though BLEU-4 gives reasonable evaluation on code generation task to some extent, we find that word error rate (WER) offers more comprehensive and sensitive evaluation for this task. It is because NMT and Transformer models appear to be better at learning the structural information of an entire code snippet than lexical information. Therefore, it requires to base evaluation measure on specialized lexicons (identifiers) and the overall structure instead of merely word grams. In addition, WER assesses one candidate with only one reference while BLEU-4 assesses with several. Since reverse engineering has only one ground truth reference, we consider WER is more appropriate for this task.

Specifically, WER measures the effectiveness of speech recognition and machine translation result, taking three common types of errors into consideration, which is computed as follows:

\begin{equation}
\mathrm{WER}=\frac{\mathrm{S}+\mathrm{I}+\mathrm{D}}{\mathrm{N}}
\end{equation}
where $S$ is the number of substitutions, $I$ is the number of insertions, $D$ is the number of deletions and $N$ is the number of words in the reference.

\begin{figure}[t]
\centering
\includegraphics[scale=0.39,clip=true]{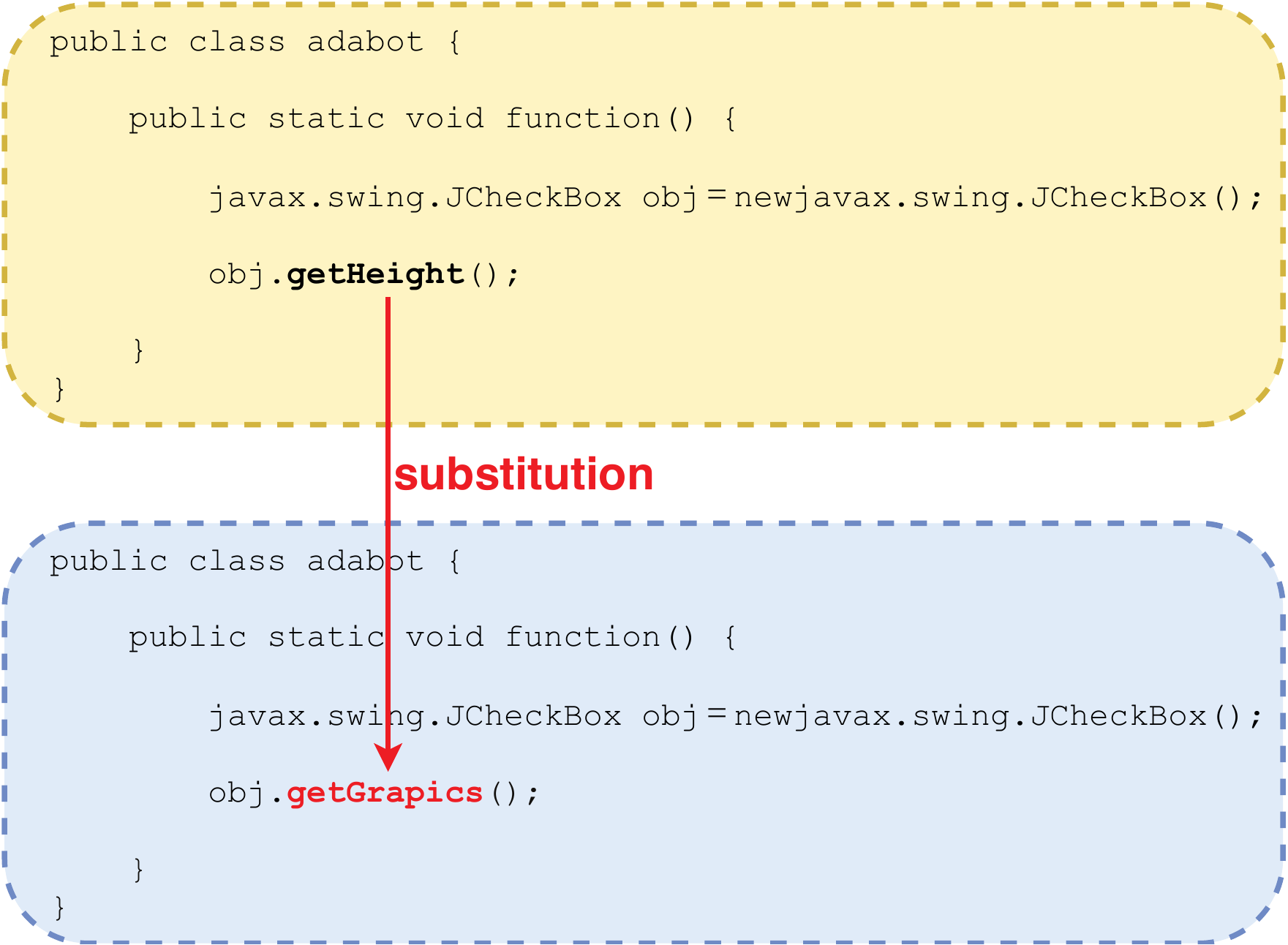} 
\caption{An example of substitution in the predicted source code}
\label{fig6}
\end{figure}

Figure \ref{fig6} shows an example of substitution in the predicted source code. Specifically, the ``getHeight'' method name in the supposed output is substituted into ``getGraphic'' which significantly changes function of the snippet and compromises the result of reverse engineering.

It is apparent that substitutions, insertions and deletions are all significant factors that may compromise the structural and lexical information in the predicted candidate, i.e. the result of reverse engineering. Thus, WER manages to offer more sensitive and rigorous evaluation on the result of reverse engineering.

\section{Experiments}

To evaluate the performance of our model, we experiment with our own corpus since there hasn't been any officially available parallel corpus of bytecode and source code. Our dataset is originally crawled from official Java 11 API offered by Oracle. Apart from evaluation of decompilation task with no error introduced, we also evaluate the performance of our model on test sets with different unit error probability to demonstrate its fault-tolerance as being a robust decompiler.

\subsection{Data Preprocessing}

Since there hasn't been any available parallel corpus for Java bytecode, mnemonic and source code, we decide to build it on our own.

\begin{table*}[t]
	\renewcommand\arraystretch{1.5}
	\centering
	\caption{Result of attention-based NMT on redundant dataset}\smallskip
	\resizebox{.99\textwidth}{!}{
		\smallskip
		\begin{tabular}{|c|c|}
			\hline
			REFERENCE&CANDIDATE \\ 
			\hline
			\makecell[c]{class javax swing JMenuItem enable\\   public static void function javax swing JMenuItem\\ obj new javax swing JMenuItemobj enable .} &\makecell[c]{ class javax swing plaf synth SynthTreeUI getClass\\  public static void function javax swing plaf synth SynthTreeUI \\obj new javax swing plaf synth SynthTreeUIobj notify .}\\
			\hline
			\makecell[c]{class javax swing JMenu isTopLevelMenu \\ public static void function javax swing JMenu\\ obj new javax swing JMenuobj isTopLevelMenu .}&\makecell[c]{class javax swing plaf synth SynthTreeUI getClass \\ public static void function javax swing plaf synth SynthTreeUI \\obj new javax swing plaf synth SynthTreeUIobj notify .}\\
			\hline
			\makecell[c]{class java lang StringBuilder reverse\\  public static void function java lang StringBuilder \\obj new java lang StringBuilderobj reverse .}&\makecell[c]{class javax swing plaf synth SynthTreeUI getClass \\ public static void function javax swing plaf synth SynthTreeUI \\obj new javax swing plaf synth SynthTreeUIobj notify .}\\
			\hline
			\makecell[c]{class javax swing tree DefaultTreeCellRenderer getRegisteredKeyStrokes\\  public static void function javax swing tree DefaultTreeCellRenderer\\ obj new javax swing tree DefaultTreeCellRendererobj getRegisteredKeyStrokes .}&\makecell[c]{class javax swing plaf synth SynthTreeUI getClass\\ public static void function javax swing plaf synth SynthTreeUI \\obj new javax swing plaf synth SynthTreeUIobj notify .}\\
			\hline
		\end{tabular}
	}
	\label{table2}
\end{table*}

\begin{figure}[t]
\centering

\subfigure[Template for static method]{
\centering
\includegraphics[scale=0.5,clip=true]{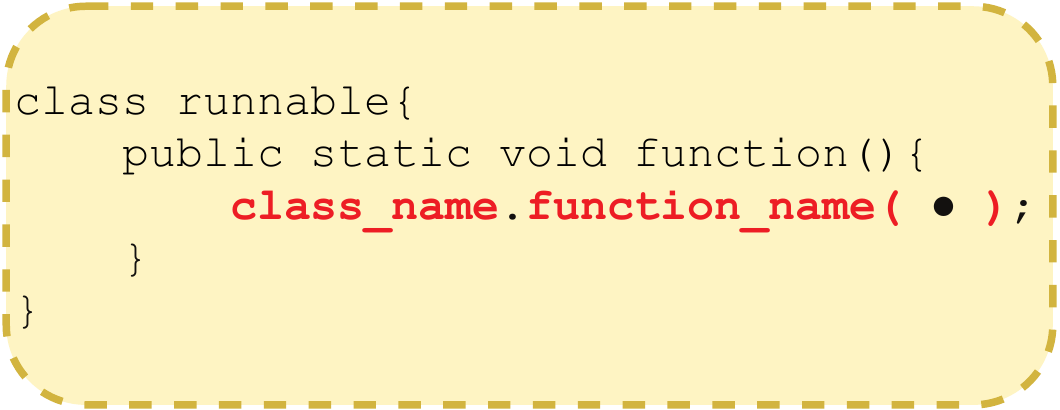} 
\label{fig7a}
}

\subfigure[Template for non-static method]{
\centering
\includegraphics[scale=0.5,clip=true]{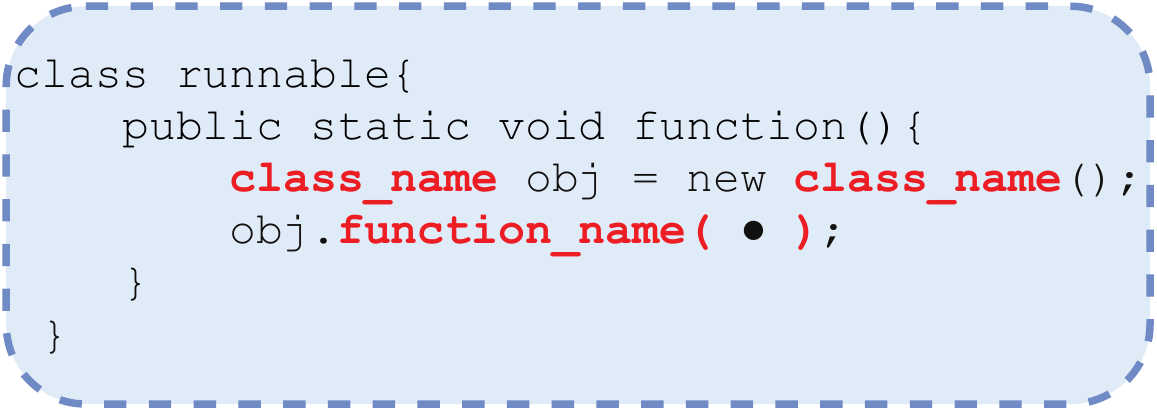} 
\label{fig7b}
}

\caption{Method template}
\label{fig7}
\end{figure}

First, we crawl all the officially available Java 11 API offered by Oracle, including name of all the classes, all the contained methods and annotations of each class. Next, in order to verify that these crawled methods are runnable (or callable), we have to match them with those rooted in our local Java libraries. To achieve this, we use the Java reflection mechanism. Concretely, we reflect all the methods (static/nonstatic) of each class and retain only those concur in both the crawled dataset and the reflected dataset. After the matching is done, we need to format them into compilable Java source code. We arbitrarily fit the static and non-static methods into two different templates, which is illustrated in Figure \ref{fig7}. Finally, in order to get our target bytecode, we use javac, the original compiler offered by Oracle Java Development Kit (JDK). We compile each of the above mentioned pre-compiled java file with it. After eliminating those cannot be compiled, we get a parallel dataset of bytecode, mnemonic and source code triple with size 18,420.

\subsection{Experiment Setup}

After preprocessing the parallel corpus. We organize them into the redundant and purified dataset. Specifically, redundant dataset consists of the original, compilable code snippets. While the purified dataset has removed a large proportion of units that represent the structural information in the bytecode and source code, only leaving those that represent the lexical information. Specifically, the identifiers.

In our experiment, we train both the attention-based NMT and base version Transformer models on redundant and purified datatsets. All batch size is set to be 16 and run for about 20 epochs until convergence for each task. It took 2.6 days to finish all the training on dual 1080 Ti GPUs. Then we use the latest checkpoint of the models to evaluate their performance on test sets with different unit error probability to evaluate their fault-tolerance.

\subsection{Results and Analysis}

We use BLEU-4 and word error rate (WER) simultaneously to evaluate the performance on each task. Not only do dual metrics offer more comprehensive evaluation, but also can we demonstrate WER as being a more suitable evaluation metric for reverse engineering task than BLEU-4.

\subsubsection{Overall Performance}

First, as Table \ref{table3} illustrated, the performance of attention-based NMT and Transformer are both pretty good and similar on the purified dataset. However, Transformer completely outweighs attention-based NMT on the redundant dataset. It is because even if attention is used as an auxiliary measure to alleviate the vanishing gradient problem which is severe when dealing with long sequences, it is only powerful enough to help it learn the structural information but not the lexical information. Therefore, we can concluded that attention mechanism is lot better in handling the unbalanced distribution of information in programming languages compared with recurrence mechanism, which makes Transformer a more suitable model for Java decompilation task than attention-based NMT.

\begin{table}[t]
	\centering
	\caption{Performance of attention-based NMT and Transformer models on the purified and redundant dataset. We evaluate each task with both BLEU-4 and WER.}\smallskip
\resizebox{.95\columnwidth}{!}{
\smallskip
	\begin{tabular}{ccc}
		\toprule  
		TASK&BLEU-4(\%)&WER(\%) \\ 
		\midrule  
		NMT purified&91.50&3.87 \\
		Transformer purified&92.30&3.48 \\
		NMT redundant&27.80&65.53 \\
     Transformer redundant&94.50&2.65 \\
		\bottomrule  
	\end{tabular}
}
\label{table3}
\end{table}

Specifically, It is apparent from row 4 in Table \ref{table3} and Table \ref{table2} that the attention-based NMT model is biased and underfitting the redundant dataset. And WER is more sensitive to its poor learning on the lexical information since WER based its evaluation on substitution, deletion and insertion of words in a sequence. Conclusively, WER is a more sensitive and informative metric for the reverse engineering task than BLEU-4.

\begin{figure*}[t]
\centering

\subfigure[Using BLEU-4 for evaluation]{
\centering
\includegraphics[width=0.48\textwidth,clip=true]{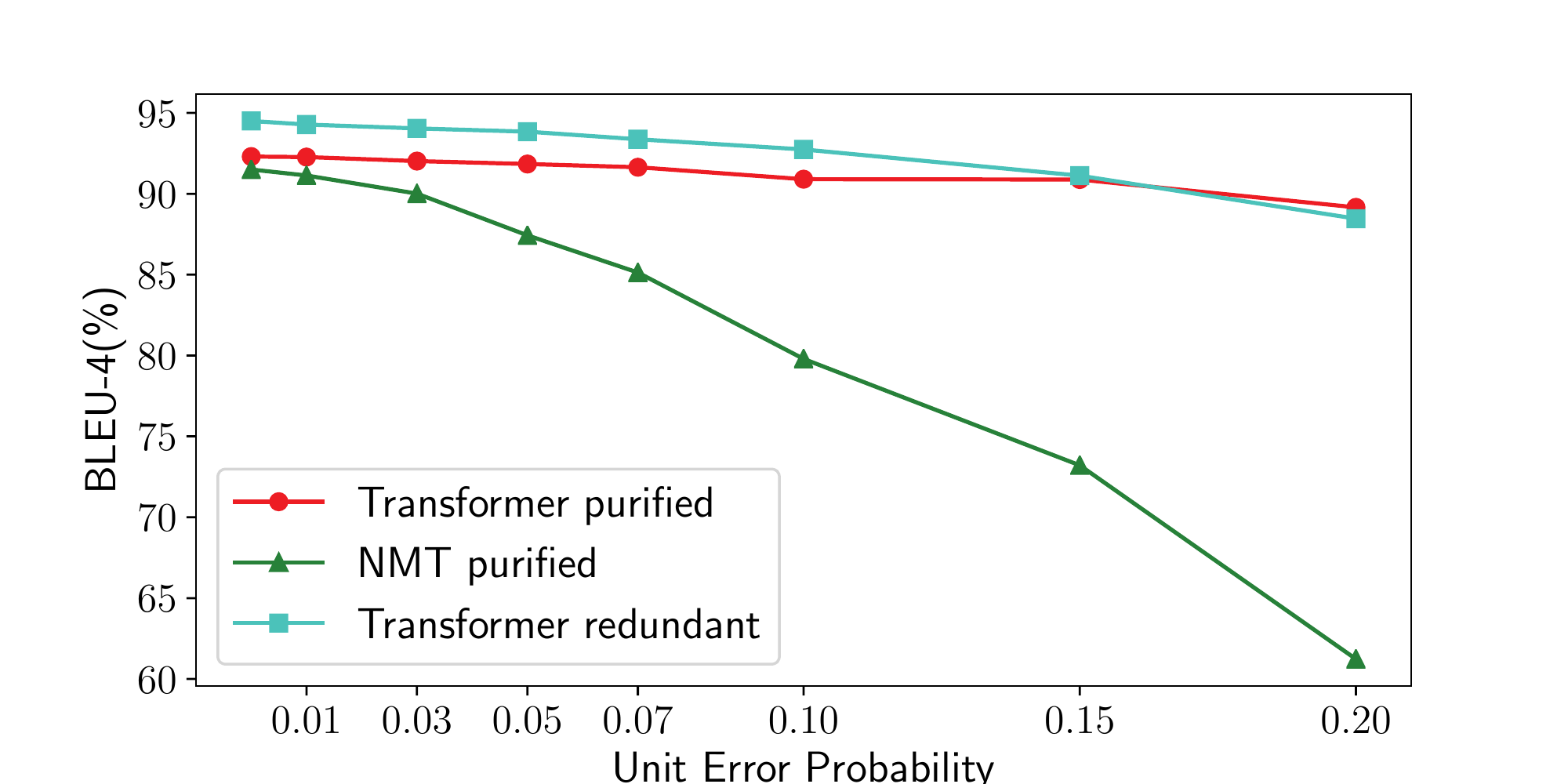} 
\label{fig8a}
}
\subfigure[Using WER for evaluation]{
\centering
\includegraphics[width=0.48\textwidth,clip=true]{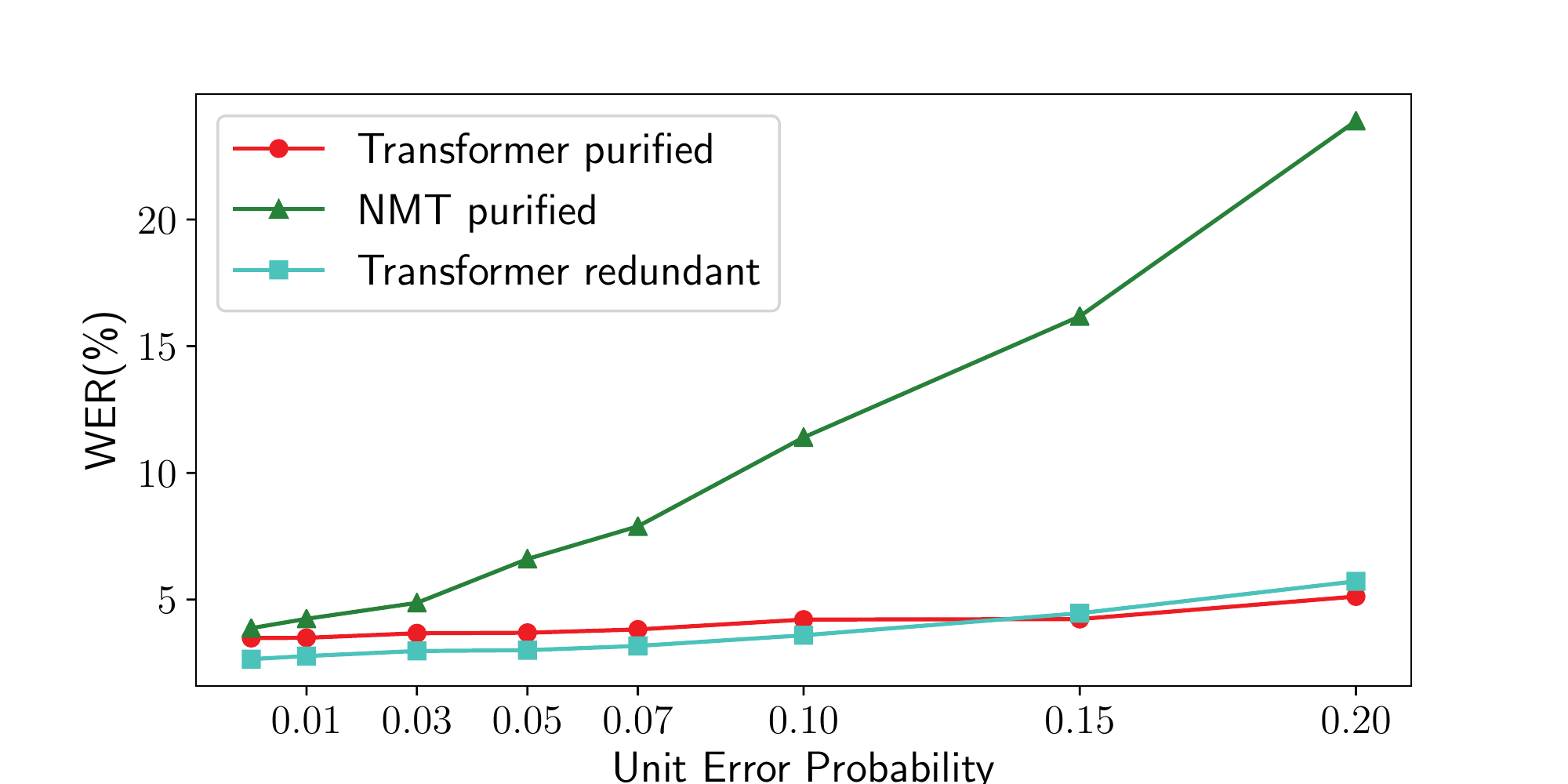} 
\label{fig8b}
}
\caption{Performance of attention-based NMT and Transformer models on purified and redundant dataset with salt-and-pepper noise introduced. The unit error probability of the noise ranges from 1\% to 20\%}
\label{fig8}
\end{figure*}

\subsubsection{Impact of Word Segmentation}

To investigate the impact of different word segmentation algorithms, we evaluate the performance of Transformer on purified dataset with different word segmentation algorithms, including space delimiter and subword model based on byte pair encoding (BPE). As Table \ref{table4} illustrated, using space delimiter for word segmentation performs better than using subword model on both metrics. The reason is because unlike natural language, programming language has a relatively small and exhaustive vocabulary. Therefore it's less likely to encounter out-of-vocabulary problem, which makes space delimiter a more suitable algorithm for word segmentation in reverse engineering task.

\begin{table}[t]
	\centering
	\caption{Impact of different word segmentation algorithms}\smallskip
\resizebox{.95\columnwidth}{!}{
\smallskip
	\begin{tabular}{ccc}
		\toprule  
		ALGORITHMS&BLEU-4(\%)&WER(\%) \\ 
		\midrule  
		space&92.30&3.48 \\
		subword model based on BPE&87.15&5.74 \\
		\bottomrule  
	\end{tabular}
}
\label{table4}
\end{table}

\subsubsection{Fault Tolerance}

In addition to evaluating the performance of our model on noise free datasets, we perform experiments on noisy dataset with salt-and-pepper noise introduced to demonstrate the fault-tolerance of our model. The unit error probability (UEP) of the salt-and-pepper noise ranges from 1\% to 20\%. We had not evaluated the performance of the attention-based NMT model on the redundant dataset since its performance on the noise free redundant dataset is already biased and not generalizable. As Figure \ref{fig8} illustrated, Transformer presents strong fault-tolerance (anti-noise ability) on both the purified and redundant datasets. Whereas, attention-based NMT model compromised quickly with the increase of UEP. Specifically, Transformer only drops 3.14\% in terms of BLEU-4 and increases 1.64\% in terms of WER on the purified dataset. And drops 6.04\% in terms of BLEU-4 and increases 3.07\% in terms of WER on the redundant dataset. However, for attention-based NMT, it is a surprising drop by 30.27\% in terms of BLEU and increase by 20.02\% in terms of WER.

The result indicates that though the performance of both attention-based NMT and Transformer models are parallel on the purified dataset with no noise introduced, Transformer presents much more stabalized and robust performance when noise is introduced in the source bytecode, whereas attention-based NMT decays rapidly with the increase of UEP. In conclusion, Transformer is not only better in handling the unbalanced distribution of information in the long sequence of source bytecode, but also much more fault-tolerant (anti-noise) compared with attention-based NMT. Therefore, it is more suitable to serve as the foundation of a robust fault-tolerant Java decompiler.

\section{Related Work}

\citeauthor{yin2018tranx} proposed an Abstract Syntax Tree (AST) based method to map natural language (NL) utterances into meaning representations (MRs) of source code; \citeauthor{hu2018deep} presented a method based on NMT model and AST to generate code comments for Java methods. Both of which investigate the relation between natural language and source code.

\citeauthor{david2019neural} proposed an approach to predict procedure names in stripped executables based on a manual encoder-decoder models; He et al., 2018 presented an approach to predict key elements of debug information in striped binaries based on probabilistic models. Both of their works investigate the relation between executables and source code.
\section{Conclusion and Future Work}

In this paper, we propose a statistical, fault-tolerant Java decompiler based on attention-based NMT and Transformer models. Specifically, using statistical models as the foundation of decompiler instead of rule-based models allow us to make the best of the fault-tolerance of statistical language models. Experimental results demonstrate that our approach not only does well in handling the unbalanced distribution of structural and lexical information in both noise free bytecode and source code, but also presents strong fault-tolerance (anti-noise ability).

For the future work, we plan to perform experiments on our model with longer, more randomized code snippets in order to further verify its robustness and get better prepared for practical use.

\bibliographystyle{aaai}
\bibliography{ref}
\end{document}